\documentclass[aps,prd,preprint,showpacs,amsmath,amssymb]{revtex4}
\usepackage{epsfig}
\newlength{\www}

\newcommand{\be}{\begin{equation}}
\newcommand{\ee}{\end{equation}}
\newcommand{\ba}{\begin{eqnarray}}
\newcommand{\ea}{\end{eqnarray}}

\newcommand{\qsl}{q\hskip-0.21cm\slash}

\newcommand{\esl}{e\hskip-0.21cm\slash}

\begin{document}

\title{\vspace{1cm}
Supersymmetry spectroscopy\\
 in stop-chargino production at LHC.
}

\author{M. Beccaria$^{a,b}$, G. Macorini$^{c, d}$,
L. Panizzi$^{c, d}$,  F.M. Renard$^e$ 
and C. Verzegnassi$^{c, d}$ \\
\vspace{0.4cm}
}

\affiliation{\small
$^a$Dipartimento di Fisica, Universit\`a di
Lecce \\
Via Arnesano, 73100 Lecce, Italy.\\
\vspace{0.2cm}
$^b$INFN, Sezione di Lecce\\
\vspace{0.2cm}
$^c$
Dipartimento di Fisica Teorica, Universit\`a di Trieste, \\
Strada Costiera
 14, Miramare (Trieste) \\
\vspace{0.2cm}
$^d$ INFN, Sezione di Trieste\\
$^e$ Laboratoire de Physique Th\'{e}orique et Astroparticules,
UMR 5207\\
Universit\'{e} Montpellier II,
 F-34095 Montpellier Cedex 5.\hspace{2.2cm}\\
\vspace{0.2cm}
}

\begin{abstract}
We consider the process of associated stop-chargino production
in the MSSM at LHC and show that, at the simplest Born level,
the production rate is dramatically sensitive to the choice 
of the benchmark points, oscillating from potentially "visible" 
maxima of the picobarn size to much smaller, hardly "visible",
values. Adopting a canonical choice of SM type CKM matrices, 
we also show that in some "visible" cases the total rate 
exhibits a possibly relevant dependence on $\tan\beta$. 

\end{abstract}


PTA/06-30\\

\maketitle

\section{Introduction}
\label{sec:intro}

\begin{figure}[tb]
\centering
\epsfig{file=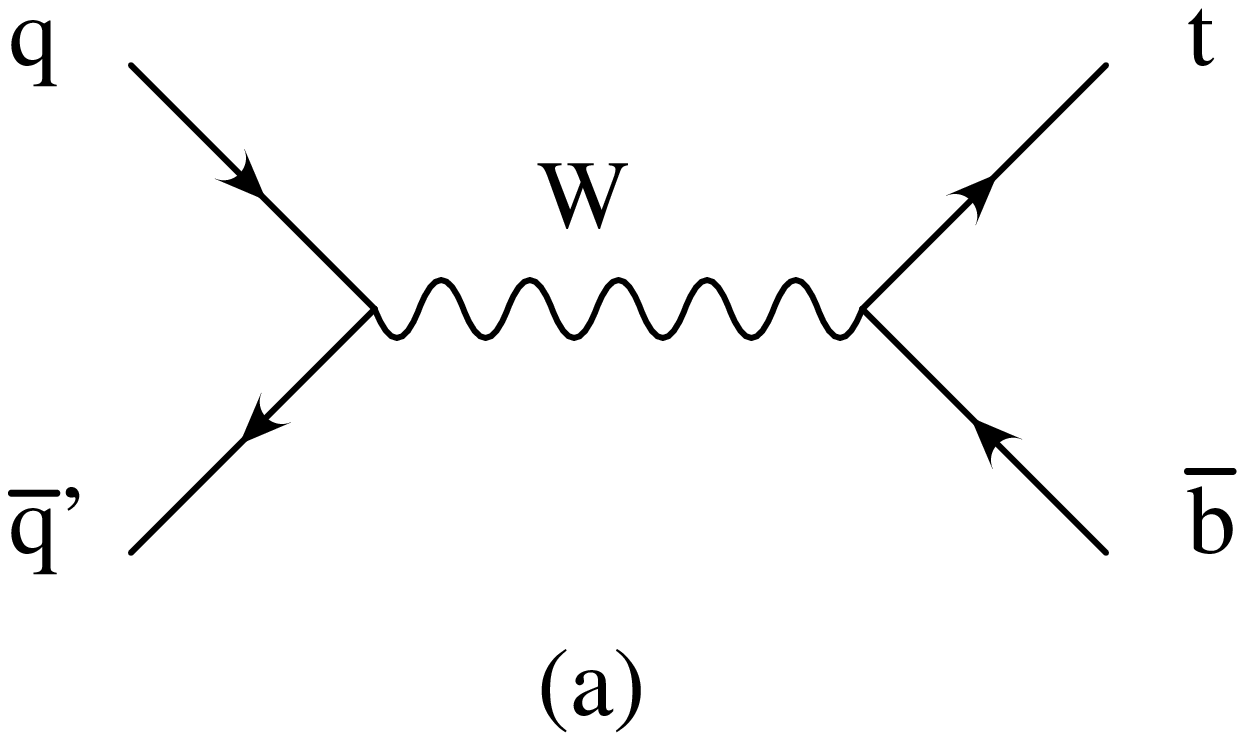, width=6cm}
\epsfig{file=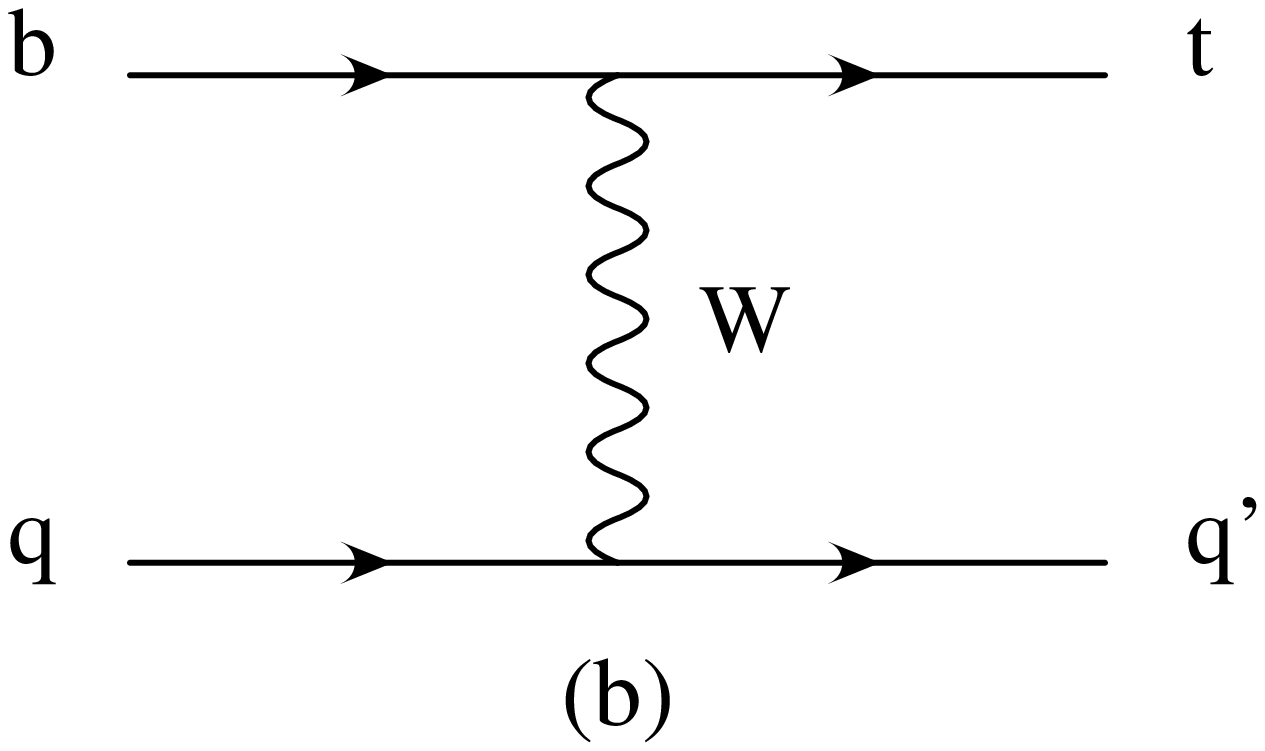, width=6cm}

\epsfig{file=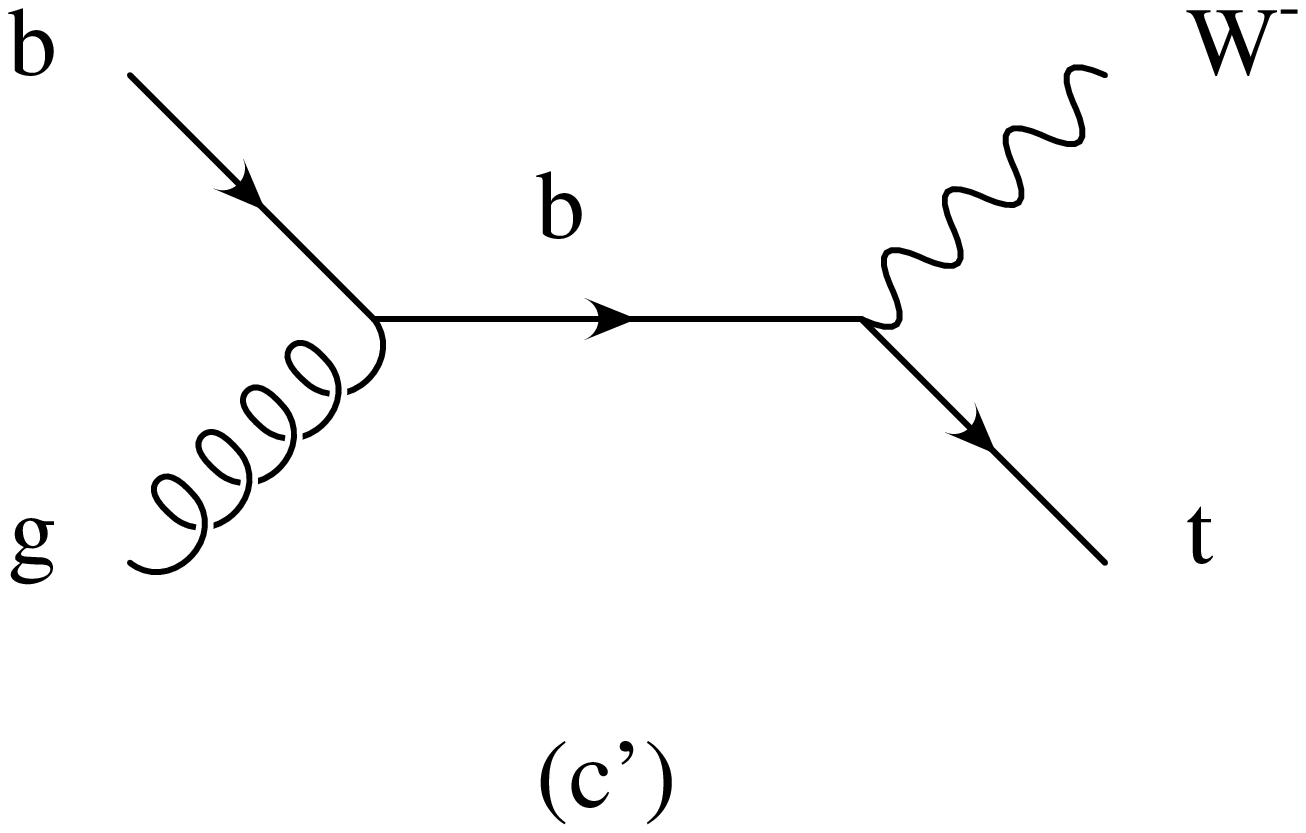, width=6cm}
\epsfig{file=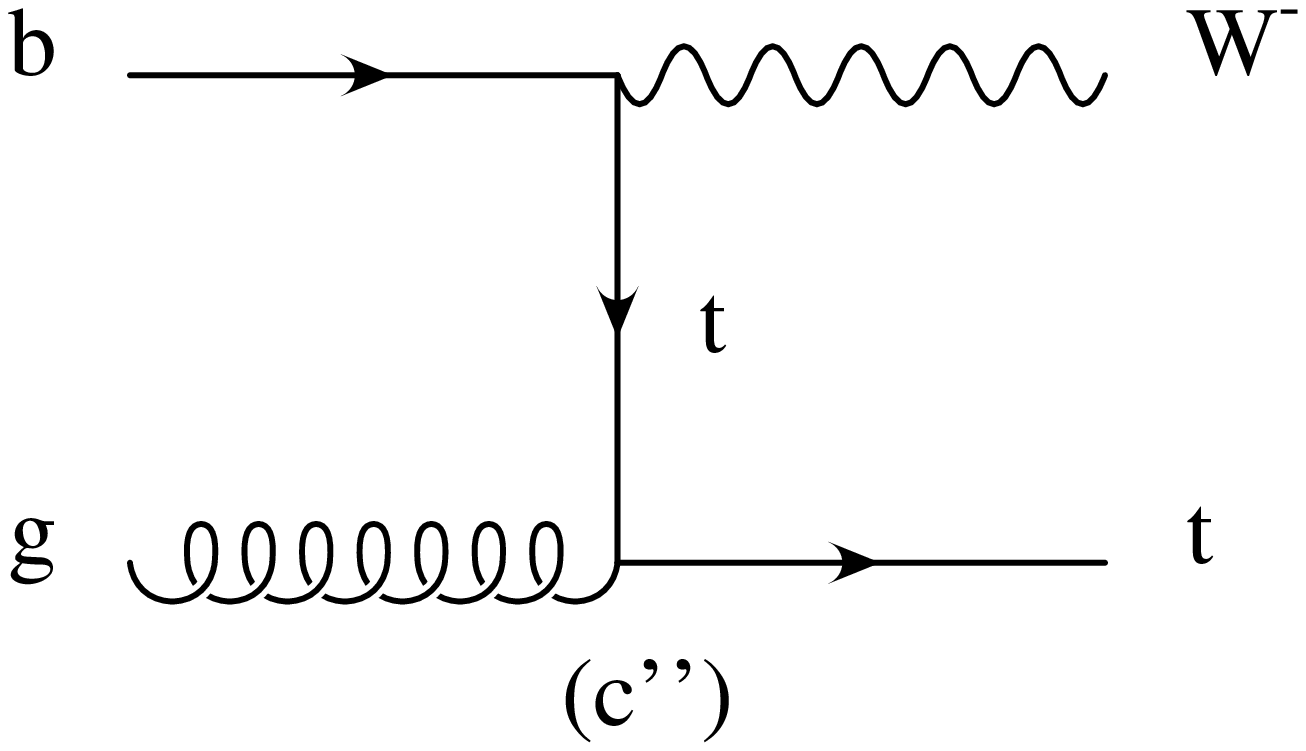, width=6cm}
\vspace{1.5cm}
\caption{Born diagrams for single top production: (a) $s$-channel, (b) $t$-channel, $(c')+(c'')$ associated production.}
\label{fig:single-top}
\end{figure}

\begin{figure}[tb]
\centering
\epsfig{file=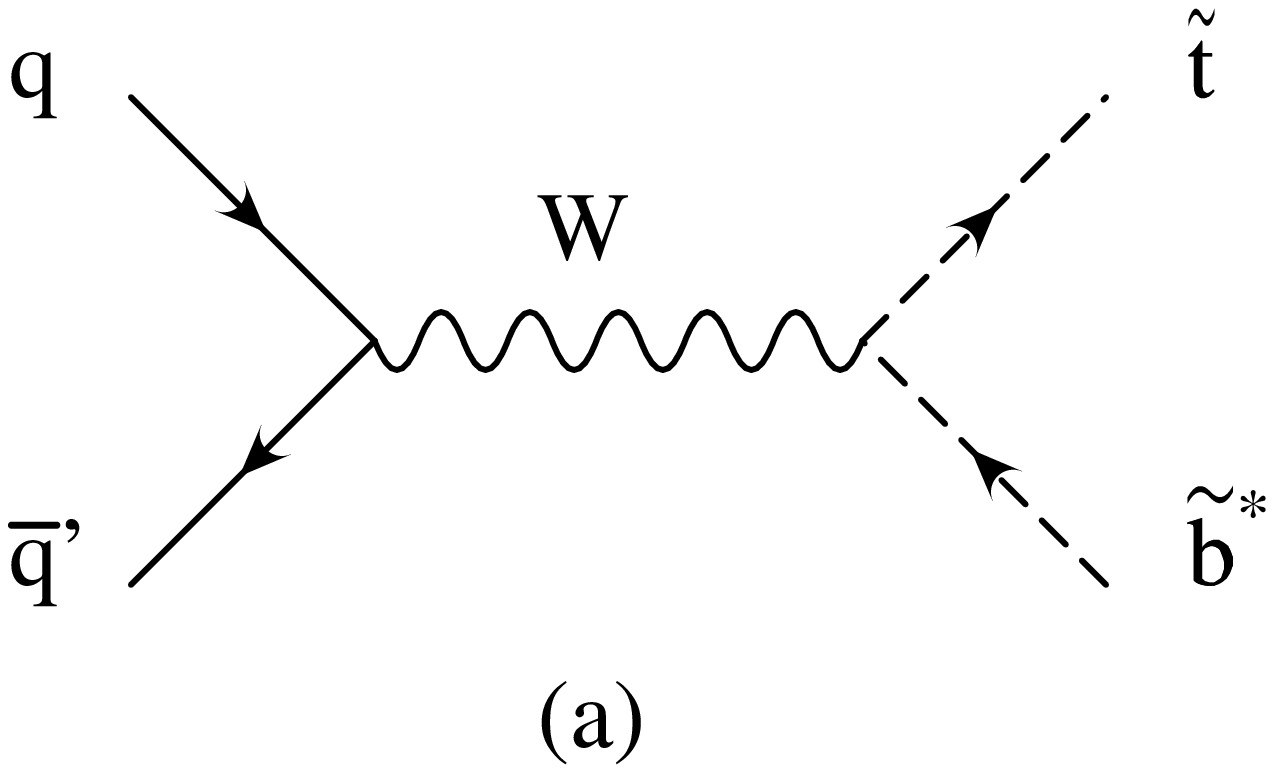, width=6cm}
\epsfig{file=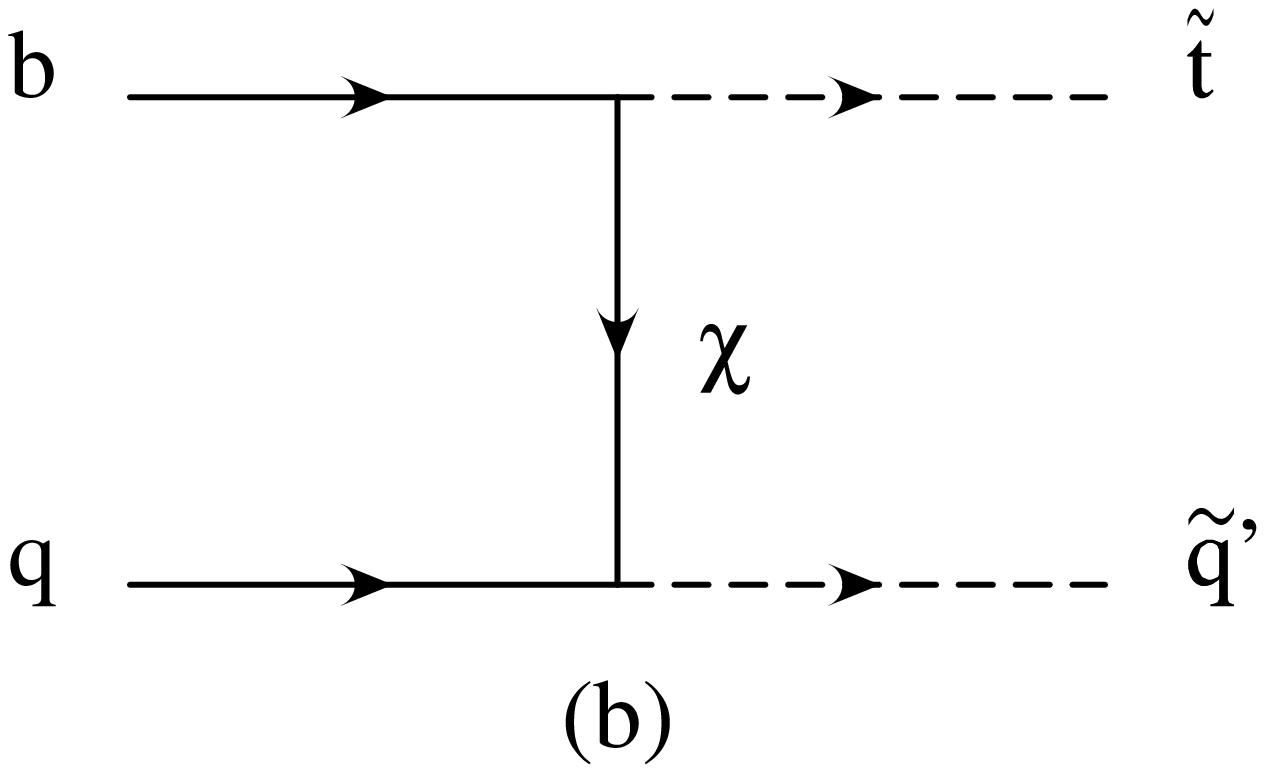, width=6cm}

\epsfig{file=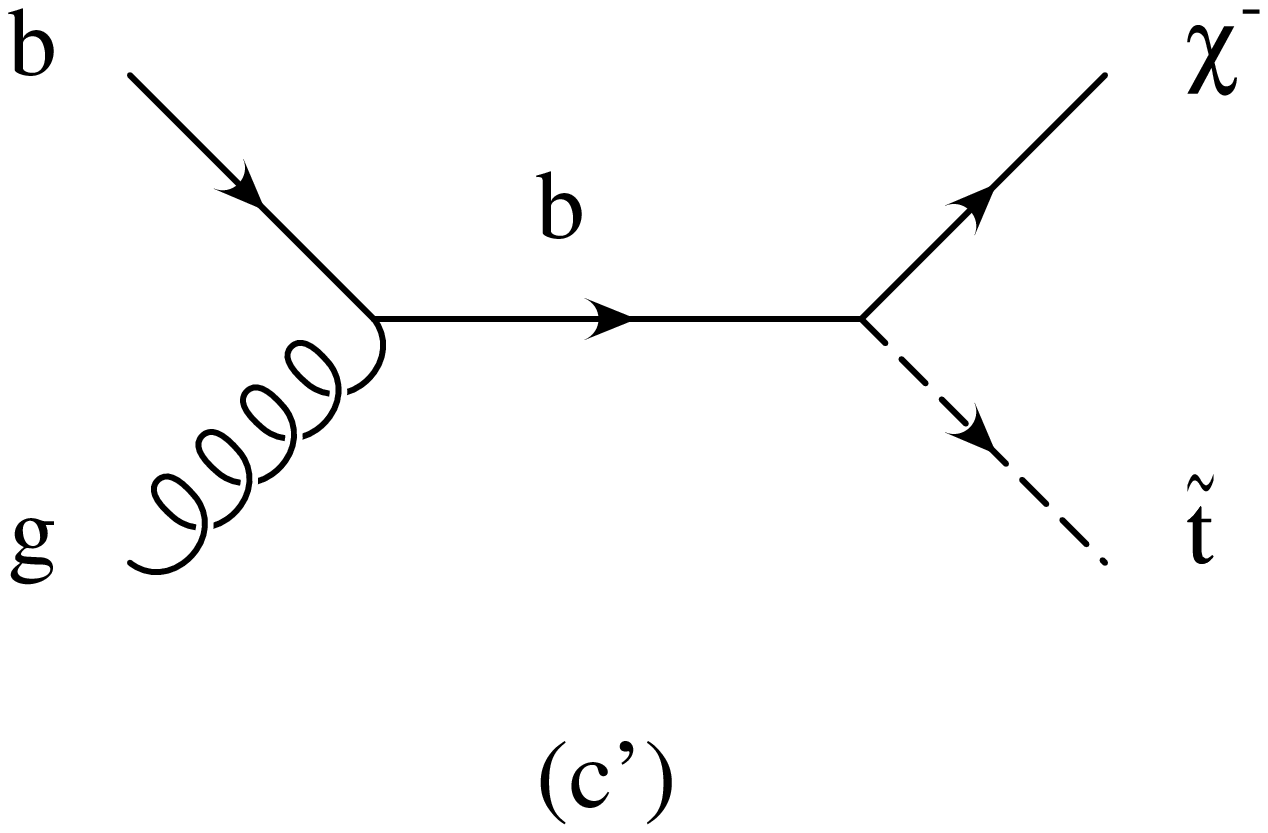, width=6cm}
\epsfig{file=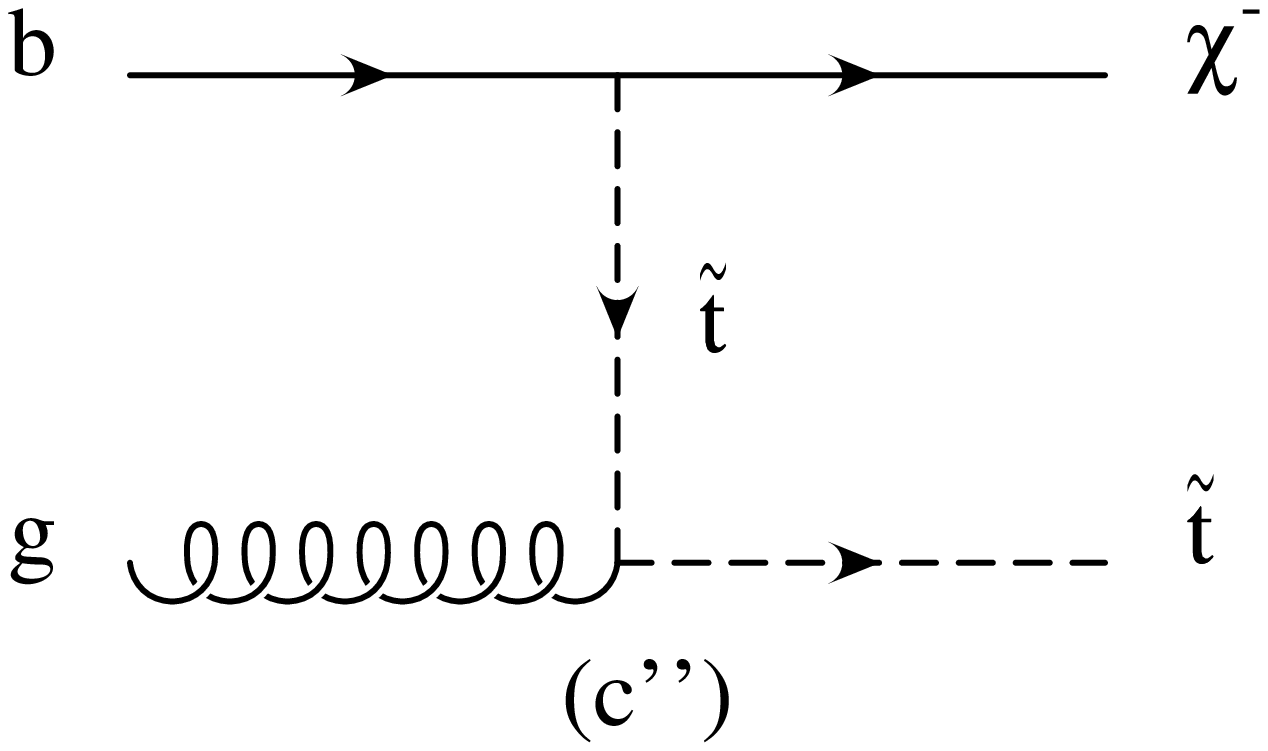, width=6cm}
\vspace{1.5cm}
\caption{Born diagrams for single stop quark production. They are in 1-1 correspondence with the diagrams of the previous Figure.}
\label{fig:single-stop}
\end{figure}

Among the various processes that will be explored at LHC, 
the single top production, represented in Fig.~(\ref{fig:single-top}), 
exhibits the interesting feature of providing 
the possibility of measuring the CKM $V_{tb}$ matrix element,
that appears already at Born level
in the related cross section, and for a recent study of
this process in the MSSM we defer to the existing literature
\cite{wt,ourpap}. In a supersymmetric scenario, a similar
property characterizes the three processes of single stop
production, represented in Fig.~(\ref{fig:single-stop}) and called in this paper,
in analogy with the single top case, $s$-channel, $t$-channel
and associated stop-chargino production. 
As stressed in a very recent work \cite{fuks},
from these processes one could
obtain a measurement of some of the CKM SUSY-SUSY matrix
elements that appear already at Born level, i.e. a test of
the usual SM-type assumptions. 
For this process, in particular the analysis of \cite{fuks} considers one
special benchmark point (SPS5) and examines the potential
effects of a deviation of the CKM matrix elements from their
SM-like values, finding sizable effects in the considered
parameter range.

Previous papers have already considered the mixed stop antisquark production~\cite{bozzi}
and the associated stop chargino production case~\cite{chinese}. In Ref.~\cite{chinese}
the process has been studied at NLO SUSY-QCD and LO electroweak level, finding a rather sizable
one-loop QCD effect, that depends very strongly on the chosen value of $\tan\beta$, that 
appears at Born level.

The calculations of~\cite{chinese} have been performed for a rather special choice of
parameters, in particular using as input a negative value $\mu = -200$ GeV. From recent experimental
analyses, the nowadays favored $\mu$ value appears to be positive if one assumes the most 
recent cosmological constraints on the dark matter relic density (see, for a very recent discussion
~\cite{Allanach:2006cc}). Given the fact that the $\tan\beta$ dependence of the stop chargino
rate appears to be very strong, we feel that it might be interesting to reconsider the analysis of~\cite{chinese}
starting from positive $\mu$ values. More specifically, the aim of this paper is that of
generalizing the analysis of Refs.~\cite{fuks,chinese} to a wider class of SUSY benchmark points with positive $\mu$.
Our main goal is that of understanding the kind of $\tan\beta$ dependence that appears in the total rate.
With this aim, we shall only consider in this preliminary paper a description given at Born level,
 and we summarize quickly the results
that we have obtained in Sec.~(II), essentially devoted to the kinematics of the process,
and in Sec.~(III), where we show the main results of our study. A short final
discussion will be given in Sec.~(IV). In Appendix~(A), we report the explicit form of the 
helicity amplitudes for the process $bg\to \widetilde{t}_a\,\chi^-_i$ at Born level.
In Appendix~(B), we give the high energy limit of the one-loop electroweak corrections.

\section{Kinematical description of the process}

The starting point is the expression of the invariant 
scattering amplitude for $bg\to \tilde{t_a}\chi^-_i$. 
In the notation of \cite{wt} we shall therefore
write (factorizing out the colour matrix elements),
the $s$-channel and $u$-channel Born contributions:

\be
A^{Born~(s)}=-\frac{g_s}{s-m^2_b}\,
\bar v_c(\chi^-_i)\,
[A^L_i(\tilde{t}_a)P_L+A^{R}_i(\tilde{t}_a)P_R]\,
(\qsl+m_b)\,\esl\, u(b)
\ee

\be
A^{Born~(u)}=-\frac{2\,g_s}{u-m^2_{\tilde{t}_a}}\,
\bar v_c(\chi^-_i)\,
[A^L_i(\tilde{t}_a)P_L+A^{R}_i(\tilde{t}_a)P_R]\,
(e.p_{\tilde{t}_a})\,u(b)
\ee
\noindent
where $e^{\mu}$ is the gluon polarization vector,
$q=p_g+p_b$, $s=(p_b+p_g)^2$, $u=(p_b-p_{\chi^-_i})^2$
and $\alpha_s={g^2_s\over4\pi}$.
The chargino states are $\chi^-_i$, $i=1,2$ whereas 
$\tilde{t}_a$, $a=1,2$ are
the physical stop states
obtained from a mixing of the chiral fields $\tilde{t}_{n}$, $n=L,R$,
such that

\be
A^{L,R}_i(\tilde{t}_a)=R_{an}A^{L,R}_i(\tilde{t}_n)
\ee
\noindent
with

\be
R_{1L}=R_{2R}=\cos\theta_t~~~R_{1R}
=-R_{2L} = \sin\theta_t
\ee
\noindent
The basic couplings
\be
A^L_i(\tilde{t}_L)=-~{e\over s_W}Z^+_{1i}~~~
A^L_i(\tilde{t}_R)={e\, m_t\over \sqrt{2}M_W\,s_W\,\sin\beta}Z^+_{2i}~~~
A^R_i(\tilde{t}_L)={e\, m_b\over 
\sqrt{2}\,M_W\,s_W\,\cos\beta}Z^{-*}_{2i}
\label{ALRi}\ee
\noindent
involve the 
chargino mixing matrices $Z^\pm_{ki}$ defined in \cite{rosiek} and controlling the 
gaugino-higgsino composition of charginos; note the
direct sensitivity on $\tan\beta$ appearing in the
higgsino components.

Starting from the previous equations and decomposing the
Dirac spinors and the gluon polarization vector
into helicity states 
$\lambda_b,\lambda_{\chi},\lambda_g$, one can easily derive 
the expression of the differential partonic cross section 
in terms of the eight possible helicity amplitudes computed in App.~(A) (we retain
the bottom mass, that cannot be neglected in a MSSM
coupling scenario):

\be
{d\sigma^{Born}\over d\cos\theta}= {\beta'\over768\pi s\beta}
\sum_{{\lambda_b,\lambda_g,\lambda_{\chi}}}
|F^{Born}_{\lambda_b,\lambda_g,\lambda_{\chi}}|^2\nonumber\\
\ee
\noindent
with $\beta={2p\over\sqrt{s}}$, $\beta'={2p'\over\sqrt{s}}$,
$p,p'$ being the initial and final c.m. momenta. The center of mass scattering angle $\theta$ is defined as the angle 
between the final stop squark and the initial bottom quark.

This partonic cross-section gets simple expressions in two limiting cases (a) at low
energy (near above the threshold $\sqrt{s_{\rm th}} = m_\chi+m_{\tilde t}$), (b) at high energy ($\sqrt{s} \gg m_\chi, m_{\tilde t}$):
\ba
(a) && \qquad \frac{d\sigma^{Born}}{d\cos\theta} = \frac{\alpha_s\,m_\chi\,\beta'}{96\,s^{3/2}\,\beta}\left[|A_i^L(\tilde t_a)|^2 + |A_i^R(\tilde t_a)|^2\right], \\ 
\nonumber \\
(b) && \qquad \frac{d\sigma^{Born}}{d\cos\theta} = \frac{\alpha_s\,\beta^{'\,3}}{96\,s\,\beta}\left[|A_i^L(\tilde t_a)|^2 + |A_i^R(\tilde t_a)|^2\right]\,\sin^2\frac{\theta}{2}.
\ea
The low energy approximation is feeded by the helicity amplitudes 
$F_{+++}$, $F_{---}$ and $F_{++-}$, $F_{--+}$ and only the $s$-channel
contribution (the $u$-channel contribution vanishes more rapidly at low energy because of an additional $\beta'$ factor coming from 
the product $e\cdot p_{\widetilde t_a}$).

The high energy approximation is only feeded by the $u$-channel amplitudes 
$F_{-++}$, $F_{+--}$. All the other ones are mass suppressed (like $m/\sqrt{s}$ or $m^2/s$) by kinematical 
factors or gauge cancellations between $s$ and $u$-channel amplitudes.
All these properties can easily be inferred from the detailed expressions listed in App.~(A).
In App.~(B), we have also written the expressions of the one-loop
electroweak corrections arising at logarithmic level from the so-called Sudakov terms~\cite{Beccaria:2003yn},
valid only in the very high energy limit.

It is remarkable that the information brought by this partonic cross section takes the  form of
\be
|A_i^L(\tilde t_a)|^2 + |A_i^R(\tilde t_a)|^2,
\ee
which, {\em e.g.} for the lighest stop quark $\widetilde t_1$, reads
\be
\cos^2\theta_t\,|A^L_i(\widetilde t_L)|^2 + \sin^2\theta_t \,|A^L_i(\widetilde t_R)|^2 + 2\,\sin\theta_t\,\cos\theta_t\,A^L_i(\widetilde t_L)\,A^L_i(\widetilde t_R) + 
\cos^2\theta_t\,|A^R_i(\widetilde t_L)|^2,
\ee
which involves the 4 parameters $\theta_t$, $\phi_{L, R}$ (that appear in the chargino mixing 
matrices~\cite{rosiek}), and $\tan\beta$.

A disentangling of these various elements could be achieved if the polarization of the produced chargino could be measured (for instance from its decay products~\cite{Choi:1998ut}).
At low energy, $\lambda_\chi = +\frac 1 2$ is feeded by $F_{+++}$ and $F_{--+}$ which only depend on $A_i^R(\widetilde t_a)$ at $\theta=0$ and only on $A_i^L(\widetilde t_a)$
at $\theta=\pi$. Conversely, $\lambda_\chi = -\frac 1 2$ is feeded by $F_{---}$ and $F_{++-}$ and only depends on $A_i^L(\widetilde t_a)$ at $\theta=0$ and only on 
$A_i^R(\widetilde t_a)$ at $\theta=\pi$.
At high energy , $\lambda_\chi = +\frac 1 2$ is only feeded by $F_{-++}$ and $A_i^L(\widetilde t_a)$, whereas 
 $\lambda_\chi = -\frac 1 2$ is produced by $F_{+--}$ and $A_i^L(\widetilde t_a)$.
In both limiting cases this would allow a good check of the stop and chargino parameters. 
For the moment we shall only concentrate in this paper on the cross-section
and its measurability at LHC.

\section{Physical observables}

The physically meaningful quantities are obtained by 
integrating over the angle with the available parton distribution functions. As a 
first observable, we considered the inclusive differential
cross section, defined as:

\ba
\label{eq:basic}
{d\sigma(PP\to \tilde{t_a}\chi^-_i+X)\over ds}&=&
{1\over S}~\int^{\cos\theta_{max}}_{\cos\theta_{min}}
d\cos\theta\, L_{bg}(\tau, \cos\theta)
{d\sigma_{bg\to  \tilde{t_a}\chi^-_i}\over d\cos\theta}(s),
\ea
\noindent
where $\tau={s\over S}$, and $L_{bg}$ is the parton process luminosity.

\be
L_{bg}(\tau, \cos\theta)=
\int^{\bar y_{max}}_{\bar y_{min}}d\bar y~ 
\left[~ b(x) g(\tau/x)+g(x)b(\tau/x)~\right]
\label{Lij}
\ee
\noindent
where S is the total $PP$ c.m. energy, and 
$i(x)$ the distributions of the parton $i$ inside the proton
with a momentum fraction,
$x={\sqrt{s\over S}}~e^{\bar y}$, related to the rapidity
$\bar y$ of the $\tilde{t_a}\chi^-_i$ system~\cite{QCDcoll}.
The parton distribution functions are the 
Heavy quark CTEQ6 set~\cite{lumi}. The rapidity and angular integrations are performed
after imposing a cut $p_T \ge 10$ GeV (see \cite{Beccaria:2004sx} for more details).

At least in the initial LHC period, a more meaningful quantity 
might be the integrated cross section. We considered here
the integration from threshold to a variable final
c.m. energy 
% referee
and allowed it to vary up to a final 
illustrative value of 2 TeV. Note that, moving to the one loop level, the Born equality between 
the final c.m. energy and the physically meaningful final invariant mass is lost. The relationship
between the two quantities can be obtained with a specific dedicated analysis analogous to the one
performed in~\cite{Beccaria:2004sx}. We expect from that reference that the difference should not be large.
This feature will be examined in details in a forthcoming paper.

In our calculation, we considered a number of benchmark
points. Our choice priviledged some special cases of
couples of points whose main difference was the value of
$\tan\beta$.  For example the two points LS1 and LS2 defined by us in~\cite{wt} exhibit an almost identical 
spectrum of physical masses, with $\tan\beta$ equal to 10 (LS1) and
50 (LS2). An analogous situation, although with slightly larger
mass differences, characterizes the two points SU1 and 
SU6~\cite{SUpoints}. To have a check of our calculation, we 
also considered the Snowmass benchmark point SPS5~\cite{SPS5} used in Ref.\cite{fuks} 
which has $\tan\beta=5$. For our purposes, 
we varied $\tan\beta$ within this point, moving to the final
(allowed) value of $\tan\beta=40$. 
The mSUGRA parameters associated with these points are reported in Tab.~(\ref{tab:benchmarks}).
\begin{table}
\begin{tabular}{||l|ccccc||}
\hline
\hline
 & $m_0$ & $m_{1/2}$ & $A_0$ & $\tan\beta$ & sign($\mu$) \\
\hline
\hline
LS1 & 300 &150 &-500 &10& +\\
LS2 & 300 &150 &-500 &50& +\\
SU1 & 70  &350 &0 &10& +\\
SU6 & 320 &375 &0 &50& +\\
SPS5 &150 &300 &-1000 &5& +\\
SPS5a &150 &300 &-1000 &15& +\\
SPS5b &150 &300 &-1000 &40& +\\
\hline
\hline
\end{tabular}
\caption{mSUGRA parameters for the benchmark points LS1, LS2, SU1, SU6, and SPS5. All masses are expressed in GeV.}
\label{tab:benchmarks}
\end{table}

\begin{figure}[tb]
\centering
\epsfig{file=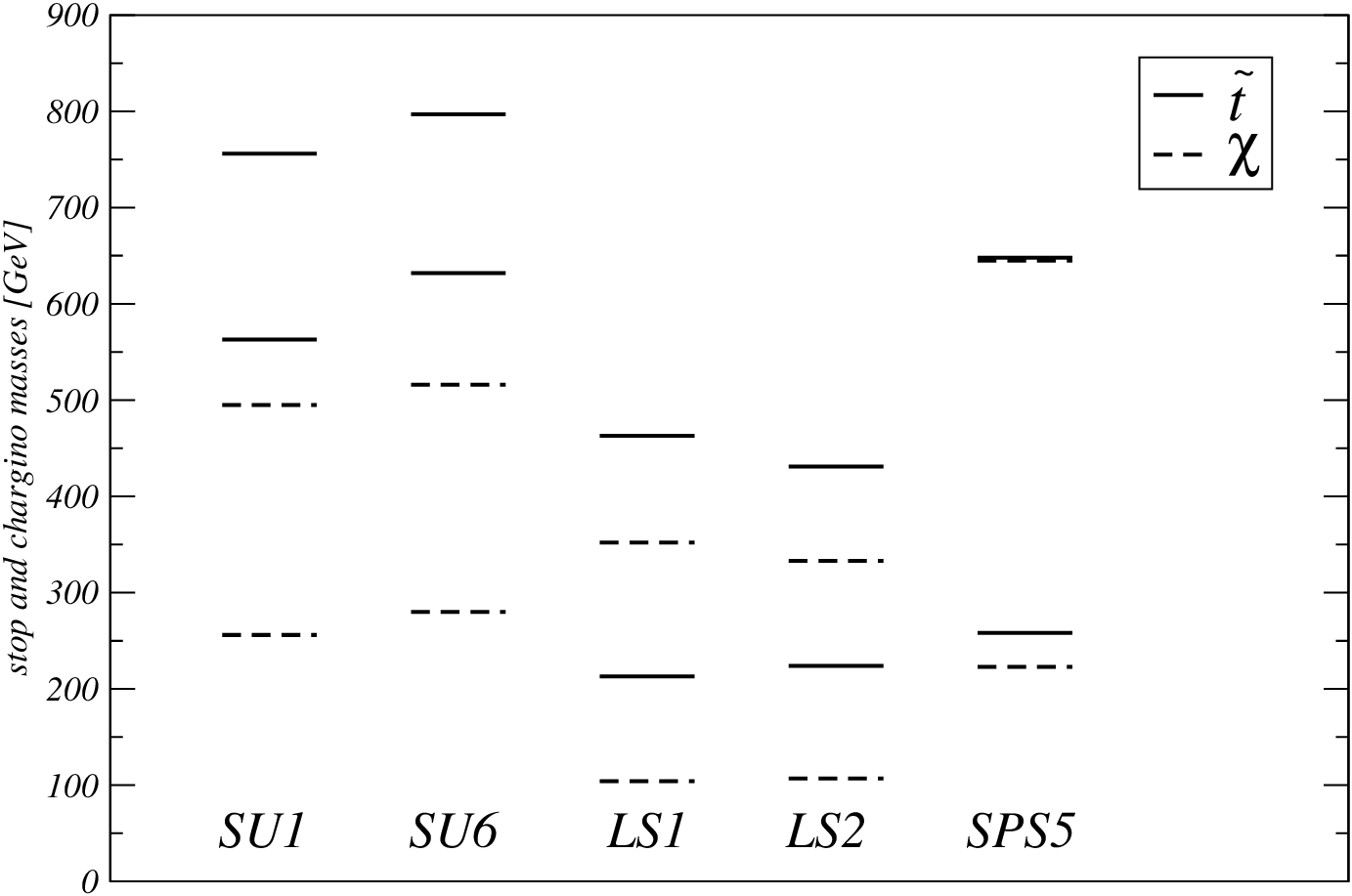, width=18cm, angle=90}
\vspace{1.5cm}
\caption{Chargino and stop masses in the considered benchmark points LS1, LS2, SU1, SU6 and SPS5.}
\label{fig:spectrum}
\end{figure}

To have a feeling of the different relevant stops, charginos 
masses in the various benchmark points, we have shown them
in Fig.~(\ref{fig:spectrum}). As one sees, the mass spectra are almost identical
in LS1 and LS2, and roughly identical in SU1 and SU6. In the
next Figures, we show the results of our calculation.

\begin{figure}[tb]
\centering
\epsfig{file=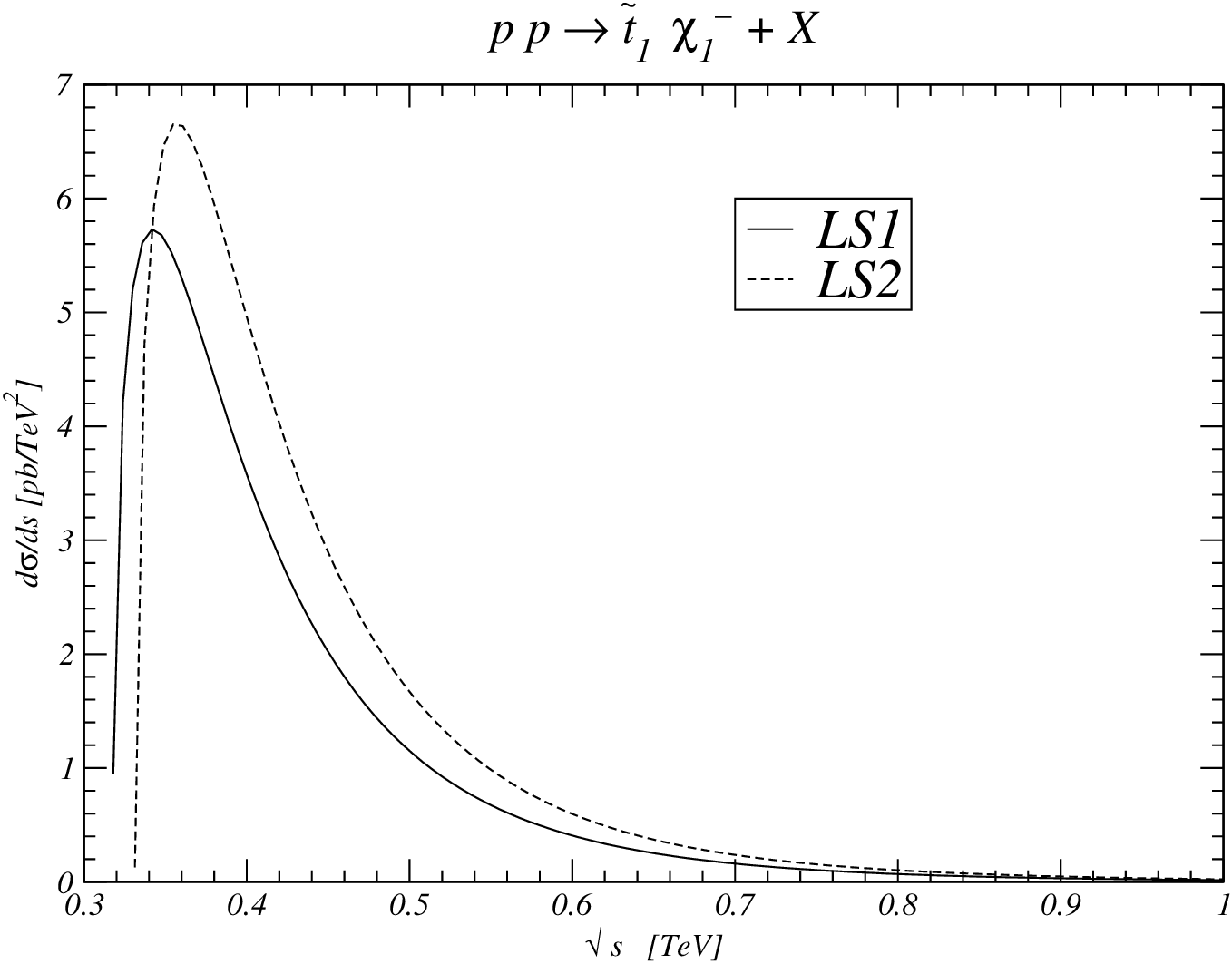, width=18cm, angle=90}
\vspace{1.5cm}
\caption{Differential distribution $d\sigma/ds$ for producing the lightest  $\widetilde{t}\,\chi$ final state at the benchmark points LS1, LS2.}
\label{fig:LS-distr}
\end{figure}

\begin{figure}[tb]
\centering
\epsfig{file=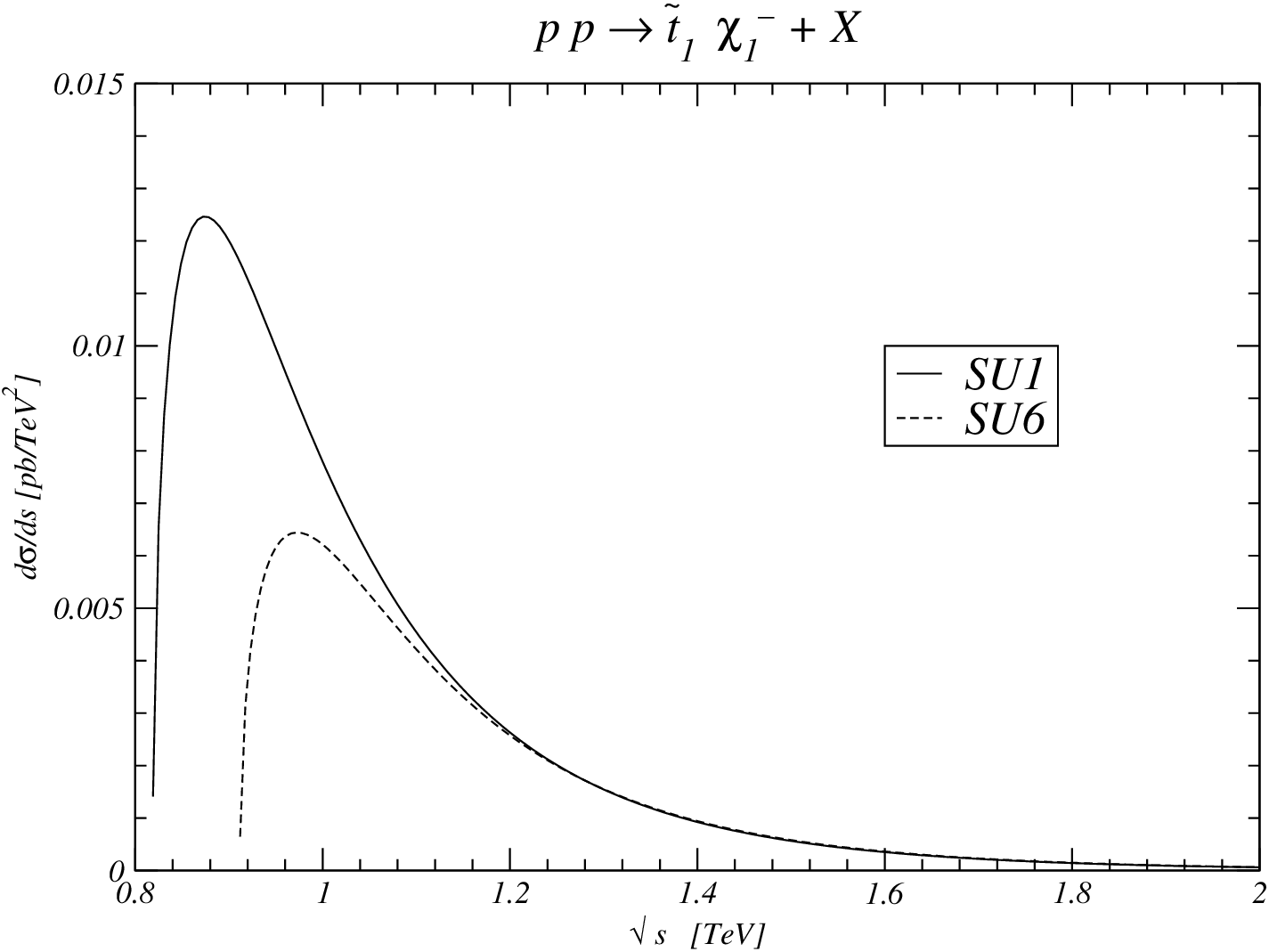, width=18cm, angle=90}
\vspace{1.5cm}
\caption{Differential distribution $d\sigma/ds$ for producing the lightest  $\widetilde{t}\,\chi$ final state at the benchmark points SU1, SU6}
\label{fig:SU-distr}
\end{figure}

\begin{figure}[tb]
\centering
\epsfig{file=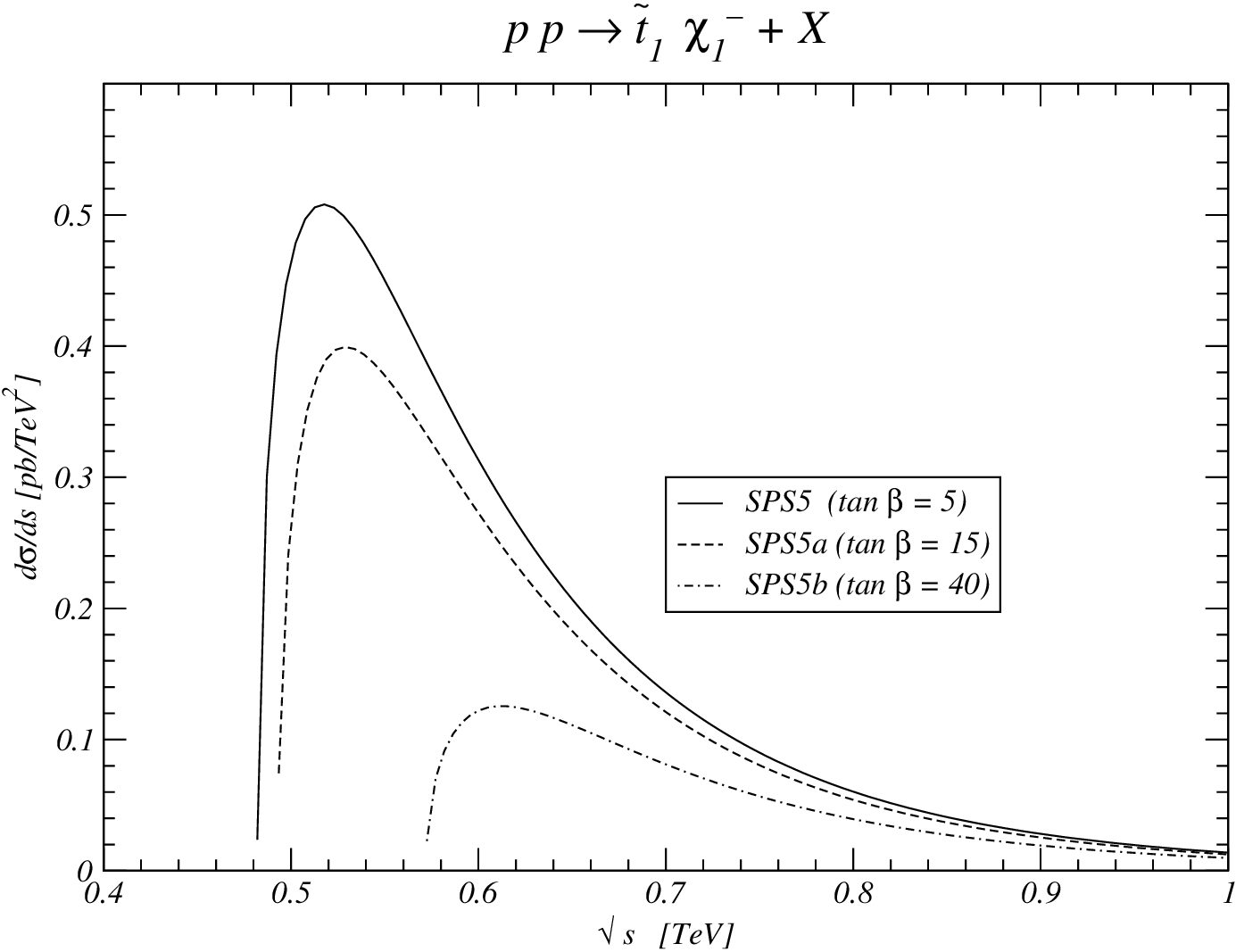, width=18cm, angle=90}
\vspace{1.5cm}
\caption{Differential distribution $d\sigma/ds$ for producing the lightest  $\widetilde{t}\,\chi$ final state at the benchmark points SPS5, SPS5a, SPS5b.}
\label{fig:SPS-distr}
\end{figure}

Figs.~(\ref{fig:LS-distr}, \ref{fig:SU-distr}, \ref{fig:SPS-distr}) 
show the inclusive differential cross sections 
for the three pairs of points. For simplicity, we only show 
the results that correspond to the lightest final pair $(\widetilde{t}, \chi)$.
We can anticipate the fact that in the three remaining 
possibilities  the distributions are greatly reduced, with
the possible exception of the combination of the lightest stop squark and 
heavier chargino in LS1, LS2 cases. As one sees, in all considered cases the common feature
is that of a sensible dependence on $\tan\beta$. 
This is due to the combination of two  quite distinct effects. 
First, the mass spectrum and, as a consequence,
the threshold for the production of a $\widetilde t\chi$ state clearly depend on $\tan\beta$. Of course, 
a high threshold implies a strong reduction of the cross-section.
This kind of effect can be observed if we compare the benchmark points SU1 and SU6 or also SPS5/SPS5a and SPS5b.
Secondly, an additional $\tan\beta$-dependence enters through the appearance of that parameter in the couplings  $A_i^{L, R}$.
This second kind of effect can be observed clearly in those cases where the variation of $\tan\beta$ does not lead to 
large changes in the threshold. Examples can be found in the comparison between the benchmark points LS1 and LS2 or SPS5 and SPS5a. 

\begin{figure}[tb]
\centering
\epsfig{file=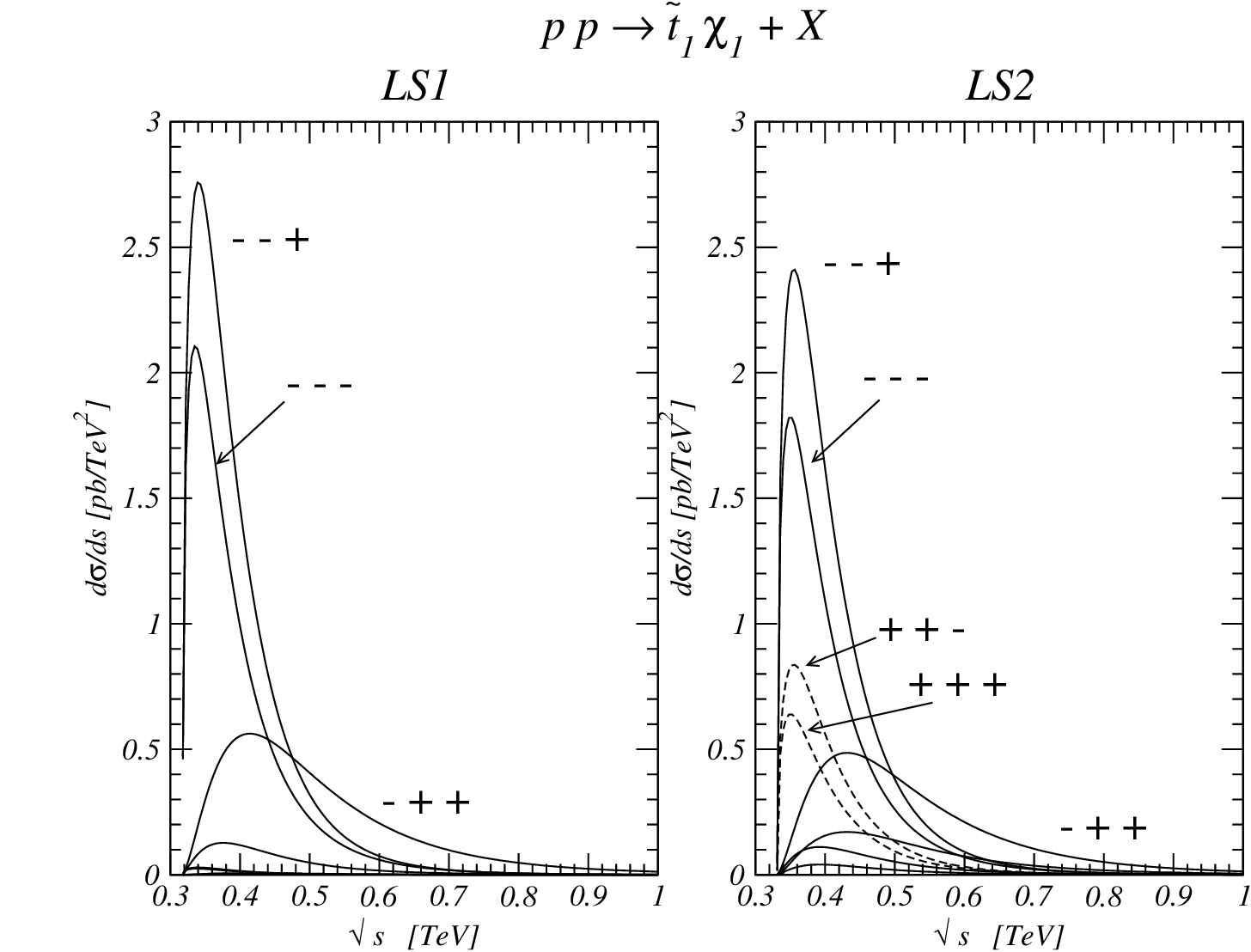, width=18cm, angle=90}
\vspace{1.5cm}
\caption{Differential distribution $d\sigma/ds$ for producing the lightest  $\widetilde{t}\,\chi$ final state at the benchmark points LS1, LS2. We show all the
separate helicity channels and label the most relevant ones. The labels show the sign of the helicities of the various involved particles in the 
order $b$ quark, gluon, chargino. An angular cut $|\cos\theta| < 0.9$ has been applied to better separate the various lines. The dashed lines in the LS2 panel 
are those helicity components which are enhanced by the large $\tan\beta$ value.}
\label{fig:LS-distr-partial}
\end{figure}

A detailed analysis of these two 
cases is worth one's while in order to understand why the cross section increases with $\tan\beta$ in the first case (LS1/LS2)
whereas it decreases in the second one (SPS5/SPS5a).
Starting with LS1/LS2, we show in Fig.~(\ref{fig:LS-distr-partial}) the distribution $d\sigma/ds$ for the various helicity components
as $\sqrt{s}$ increases. In the peak region, there are two dominant contributions 
with the specific (sign of) helicity combinations $--+$ and $---$ whereas at higher energy, only the $-++$ term survives.
When $\tan\beta$ is increased, the dominant channels tend to decrease slightly. They receive a mixed contribution composed of 
a dominant gaugino coupling and a smaller higgsino one. The relative signs of the combinations are responsible for the 
decrease of this term when  $\tan\beta$ is increased and the higgsino contribution grows. On the other hand, there are two  
channels with helicities $++-$ and $+++$ which are purely of higgsino origin. These are quite small in LS1 but sizable in LS2.
They are responsible for the increase in the total $d\sigma/ds$. We do not show a similar figure for the case SPS5/SPS5a,
but the mechanism is quite similar. The only difference is that now the $++-$ and $+++$ amplitudes are much smaller
due to a combination of various effects (different $\tan\beta$ and mixing matrix elements). When $\tan\beta$ is increased from 5 to 15,
the rise of these amplitudes is unable to invert the (negative) trend due to the leading amplitudes.

To make a more realistic analysis, 
we show in the next Figs.~(\ref{fig:LS-sigma}, \ref{fig:SU-sigma}, \ref{fig:SPS-sigma})  the values
of the integrated cross sections. From their inspection, two main conclusions can
be, in our opinion, derived. The first one concerns the 
magnitude of the various rates. We assume, for simplicity,
that a value of one picobarn for the rate corresponds,
roughly and for the expected luminosity, to ten thousands
events per year. In this sense, we consider it as a 
reasonable experimentally meaningful limit, leaving aside
in this qualitative discussion identification details. For
rate values drastically below the picobarn size, the process
might still be "visible" but hardly exploitable in our opinion 
for a meaningful parameter analysis.

Keeping in mind our qualitative classification, we see that
the rates of the three point couples oscillate from a maximum
of the picobarn size (LS1, LS2) to a minimum
of $\simeq 10^{-2}$ picobarn (SU1, SU6), passing through an
intermediate stage of $\simeq 10^{-1}$ picobarn (SPS5) (in
fact, all our results should be multiplied by a factor of
two, to keep into account the conjugate state 
$\chi^+\tilde{t}^*$, produced by an initial $\bar b\, g$
pair, whose rate is essentially identical with that of
$\chi^-\tilde{t}$). As one sees, the rate variations from
one pair to another one are of one order of magnitude. 
We should also say as a check of our calculations, that our 
results for SPS5 reproduce essentially the corresponding 
SM type one (defined "plain") of Ref.\cite{fuks}.

The second main comment is that, in all three cases, the
variations of the rate with $\tan\beta$ are not small.
More precisely, they are of the relative thirty percent
size for LS1, LS2 and of the relative sixty percent size
for SU1, SU6. In the case of SPS5, the variation with 
$\tan\beta$ is the most drastic. Moving from $\tan\beta=5$
to $\tan\beta=40$ would change the rate by a factor of 3
that would hardly escape an experimental detection.

We remark that this rather strong $\tan\beta$ dependence
is observable despite the gaugino character of the lightest chargino, a general feature of mSUGRA 
benchmark points with a light stop. Moving onward to more general symmetry breaking schemes,
any point characterized by a light stop and an higgsino-like lightest chargino would enhance 
the discussed sensitivity.
This would be precisely the case of~\cite{chinese}, since their choice of the parameters
selects, as one can easily check, a lightest chargino with a sizable higgsino component.

\begin{figure}[tb]
\centering
\epsfig{file=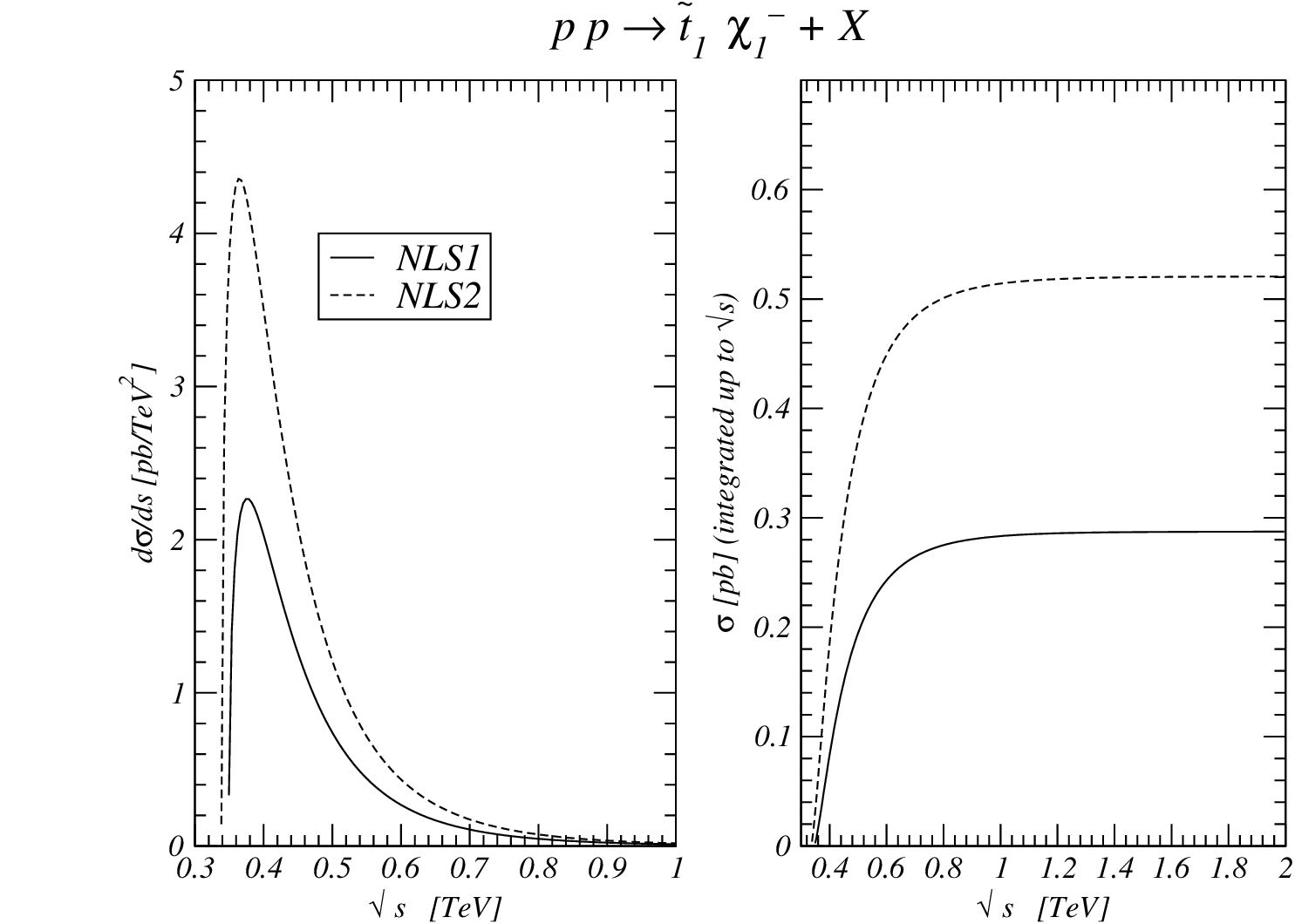, width=18cm, angle=90}
\vspace{1.5cm}
\caption{Differential distribution $d\sigma/ds$ and integrated cross section from threshold up to $\sqrt{s}$ 
for the production of the  lightest  $\widetilde{t}\,\chi$ final state. The benchmark points are NLS1, NLS2, both with $\mu<0$.}
\label{fig:NegLS}
\end{figure}

\begin{figure}[tb]
\centering
\epsfig{file=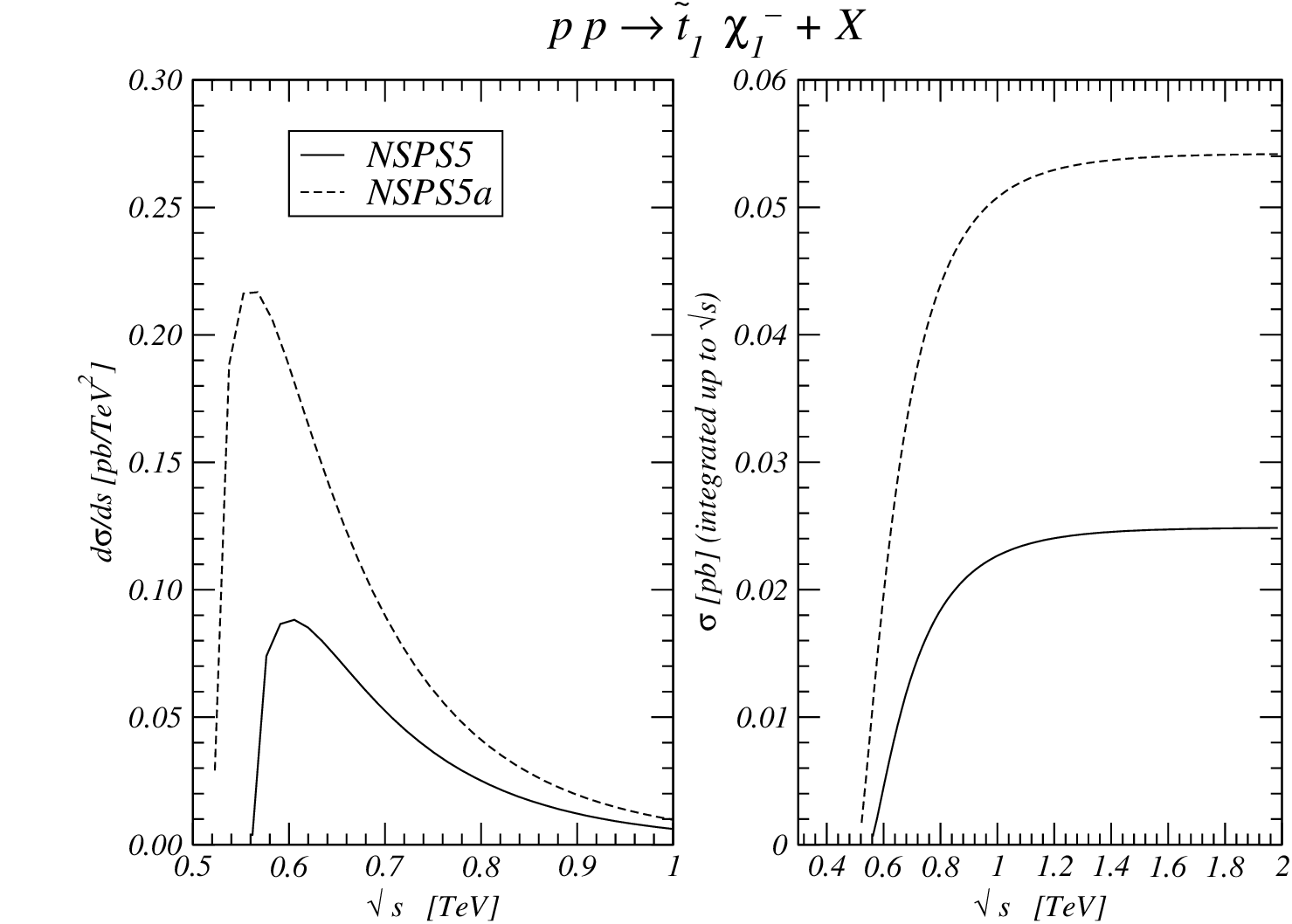, width=18cm, angle=90}
\vspace{1.5cm}
\caption{Differential distribution $d\sigma/ds$ and integrated cross section from threshold up to $\sqrt{s}$ 
for the production of the  lightest  $\widetilde{t}\,\chi$ final state. The benchmark points are NSPS5, NSPS5a, both with $\mu<0$.}
\label{fig:NegSPS5}
\end{figure}

A final possibly interesting question that we try to face is the dependence of the rate on 
the assumed sign of the $\mu$ parameter (this could become relevant only 
if the existing cosmological constraint were removed). To give a quantitative example of the 
potential effects of a change in the sign of $\mu$, we have redone some of our calculations
changing the sign of the $\mu$ value. Figs.~(\ref{fig:NegLS}-\ref{fig:NegSPS5}) show the results
in the two ``more observable'' cases LS1/LS2 and SPS5/SPS5a. The associated benchmark points with 
{\em negative} $\mu<0$ (and the same $|\mu|$) are defined  NLS1, NLS2, NSPS5, and NSPS5a. As one sees, the $\tan\beta$ dependence is strongly
enhanced in the two cases, increasing by almost a factor 2 with respect to the positive $\mu$ analysis,
while the size of the total rate is only slightly decreased, at least for large $\tan\beta$.
Note that even with a negative $\mu$ the lightest chargino is still of essentially gaugino type
due to the natural $|\mu|\gg M_2$ hierarchy, contrary to the case of~\cite{chinese}. 

\section{Conclusions}

In conclusion, we have seen that the process of stop-chargino
production appears, in a light stop-chargino scenario,
 to be a possible promising candidate for
a ``spectroscopic'' test of different SUSY scenarios and also, possibly, for a 
measurement of $\tan\beta$. 
% referee
In this respect, it would represent an alternative possibility to those offered by more conventional and direct 
measurements in the Higgs sector. For instance, at LHC, $\tan\beta$ can be determined from a study of the production 
process $gg\to b\, \bar b\, H/A/h$ which is the dominant Higgs boson production process at large $\tan\beta$~\cite{Dittmaier:2003ej}.
From an analysis of the subsequent Higgs decays (mainly $H/A/h\to \tau\tau \to jj+X$ or $H/A/h\to \tau\tau \to \ell j+X$)
performed at 30 fb${}^{-1}$ it is possible to determine $\tan\beta$ with a statistical error $4-25$ \% and a systematic
error $\le 12$ \% depending on the signal significance~\cite{Hashemi:2005qd}.

To complete this preliminary study, two main
analyses are still missing, The first one is a realistic
experimental discussion of the expected errors. 
To our knowledge, preliminary experimental analyses of light stop-antistop
pairs production have been very recently provided~\cite{Iris}. In our opinion, their extension to the 
stop-chargino case could be interesting.
The second one is a complete NLO theoretical calculation, which would be
justified by the presence of a large $\tan\beta$ dependence
at Born level. 
For what concerns the NLO QCD analysis for a more general set of input parameters, in particular with $\mu>0$, 
we think that the analysis of~\cite{chinese} should be redone, but an essential point, in our opinion,
would be the additional and combined calculation of the one loop electroweak effects, since a priori
the $\tan\beta$ dependence might be sensibly modified at this level, particularly in a light 
stop chargino scenario where sizable electroweak logarithmic effects of Sudakov kind shown in App.~(B)
might arise from one-loop diagrams. Our group is already proceeding in the complete one-loop
electroweak calculation.

\appendix
\section{Born level helicity amplitudes}

The Born level helicity amplitude is 
\be
F_{\lambda_b\,\lambda_g\,\lambda_\chi} = \mathop{\sum_{\eta=L, R}}_{k=1,2} 
\, N^\eta_k\,{\cal H}^\eta_{k,\ \lambda_b\,\lambda_g\,\lambda_\chi},
\ee
where 
\be
N^\eta_1 = -g_s\,{A^{\eta}_i(\tilde{t}_a)\over s-m^2_b}, \qquad
N^\eta_2 = 2\,g_s\,{A^{\eta}_i(\tilde{t}_a)\over u-m^2_{\tilde{t}_a}},
\ee
and
\ba
{\cal H}^\eta_{1, +++} &=& -{pR\over\sqrt{2}}(1+r_b)(1+\eta)(1-r_{\chi})\cos{\theta\over2} \\
{\cal H}^\eta_{1, ++-} &=& -{pR\over\sqrt{2}}(1+r_b)(1+\eta)(1+r_{\chi})\sin{\theta\over2} \\
{\cal H}^\eta_{1, --+}&=& {pR\over\sqrt{2}}(1+r_b)(1-\eta)(1+r_{\chi})\sin{\theta\over2} \\
{\cal H}^\eta_{1, ---}&=& -{pR\over\sqrt{2}}(1+r_b)(1-\eta)(1-r_{\chi})\cos{\theta\over2} 
\ea
\ba
{\cal H}^\eta_{2, +++}&=& -{p'R\sin\theta\over2\sqrt{2}}(1+\eta r_b-r_{\chi}(\eta+r_b))\sin{\theta\over2} \\
{\cal H}^\eta_{2, +-+}&=& {p'R\sin\theta\over2\sqrt{2}}(1+\eta r_b-r_{\chi}(\eta+r_b))\sin{\theta\over2}\\
{\cal H}^\eta_{2, ++-}&=& {p'R\sin\theta\over2\sqrt{2}}(1+\eta r_b+r_{\chi}(\eta+r_b))\cos{\theta\over2}\\
{\cal H}^\eta_{2, +--}&=& -{p'R\sin\theta\over2\sqrt{2}}(1+\eta r_b+r_{\chi}(\eta+r_b))\cos{\theta\over2}\\
{\cal H}^\eta_{2, -++}&=& {p'R\sin\theta\over2\sqrt{2}}(1-\eta r_b-r_{\chi}(\eta-r_b))\cos{\theta\over2}\\
{\cal H}^\eta_{2, --+}&=& -{p'R\sin\theta\over2\sqrt{2}}(1-\eta r_b-r_{\chi}(\eta-r_b))\cos{\theta\over2}\\
{\cal H}^\eta_{2, -+-}&=& {p'R\sin\theta\over2\sqrt{2}}(1-\eta r_b+r_{\chi}(\eta-r_b)) \sin{\theta\over2}\\
{\cal H}^\eta_{2, ---}&=& -{p'R\sin\theta\over2\sqrt{2}}(1-\eta r_b+r_{\chi}(\eta-r_b)) \sin{\theta\over2}
\ea
The kinematical parameters $R$, $r_b$, and $r_\chi$ appearing in the above expressions 
are defined as
\ba
R &=& \sqrt{(E_b+m_b)(E_{\chi}+m_{\chi})}, \\
r_b &=& {p\over E_b+m_b}, \\
r_\chi &=& {p'\over E_{\chi}+m_{\chi}},
\ea
where $E_b$ and $E_\chi$ are the b quark and chargino c.m. energies.

\section{One-loop electroweak corrections at logarithmic level}

From the general rules given in~\cite{Beccaria:2003yn} we can already give the expressions of these
corrections due to the so-called Sudakov terms. These expressions should only be valid in the 
domain $\sqrt{s}\gg m_{\widetilde t}, m_\chi$. At low energy, a dedicated complete one-loop 
calculation is necessary and is under way.

\ba
F^{Sudakov}_{-++}(\widetilde{t}_L) &=& F^{Born}_{-++}(\widetilde{t}_L) \,\frac{\alpha}{4\pi}
\left\{\frac{1+26\,c_W^2}{36\,s_W^2\,c_W^2}\left(2\,\log\frac{s}{M_W^2}-\log^2\frac{s}{M_W^2}\right) + \right. \\
&& \left. -\frac{1}{s_W^2}\,\log\frac{s}{M_W^2}\left[
2\,\log\frac{-u}{s}+\frac{1-10\,c_W^2}{18\,s_W^2\,c_W^2}\,\log\frac{-t}{s}
\right] + \right. \nonumber \\
&& \left. -\log\frac{s}{M_W^2}\left[
\frac{m_t^2 (1+\cot^2\beta)}{2 s_W^2 M_W^2}+ \frac{m_b^2 (1+\tan^2\beta)}{2 s_W^2 M_W^2}
\right]
\right\} \nonumber \\ \nonumber \\
F^{Sudakov}_{-++}(\widetilde{t}_R) &=& F^{Born}_{-++}(\widetilde{t}_R) \,\frac{\alpha}{4\pi}
\left\{-\frac{13+14\,c_W^2}{36\,s_W^2\,c_W^2}\,\log^2\frac{s}{M_W^2} + \right. \\
&& \left. -\frac{1}{6 c_W^2}\,\log\frac{s}{M_W^2}\left[
\frac{4}{3}\,\log\frac{-t}{s}-\frac{1-10\,c_W^2}{s_W^2}\,\log\frac{-u}{s}
\right]  \right\} \nonumber \\ \nonumber \\
F^{Sudakov}_{+--}(\widetilde{t}_L) &=& F^{Born}_{+--}(\widetilde{t}_L) \,\frac{\alpha}{4\pi}
\left\{-\frac{7+20\,c_W^2}{36\,s_W^2\,c_W^2}\,\log^2\frac{s}{M_W^2} + \right. \\
&& \left. -\frac{1}{3 c_W^2}\,\log\frac{s}{M_W^2}\left[
\log\frac{-u}{s}-\frac{1}{3}\,\log\frac{-t}{s}
\right]  \right\} \nonumber 
\ea

\newpage

\newpage

\newpage

\newpage

\newpage

\newpage

\newpage
\begin{figure}[tb]
\centering
\epsfig{file=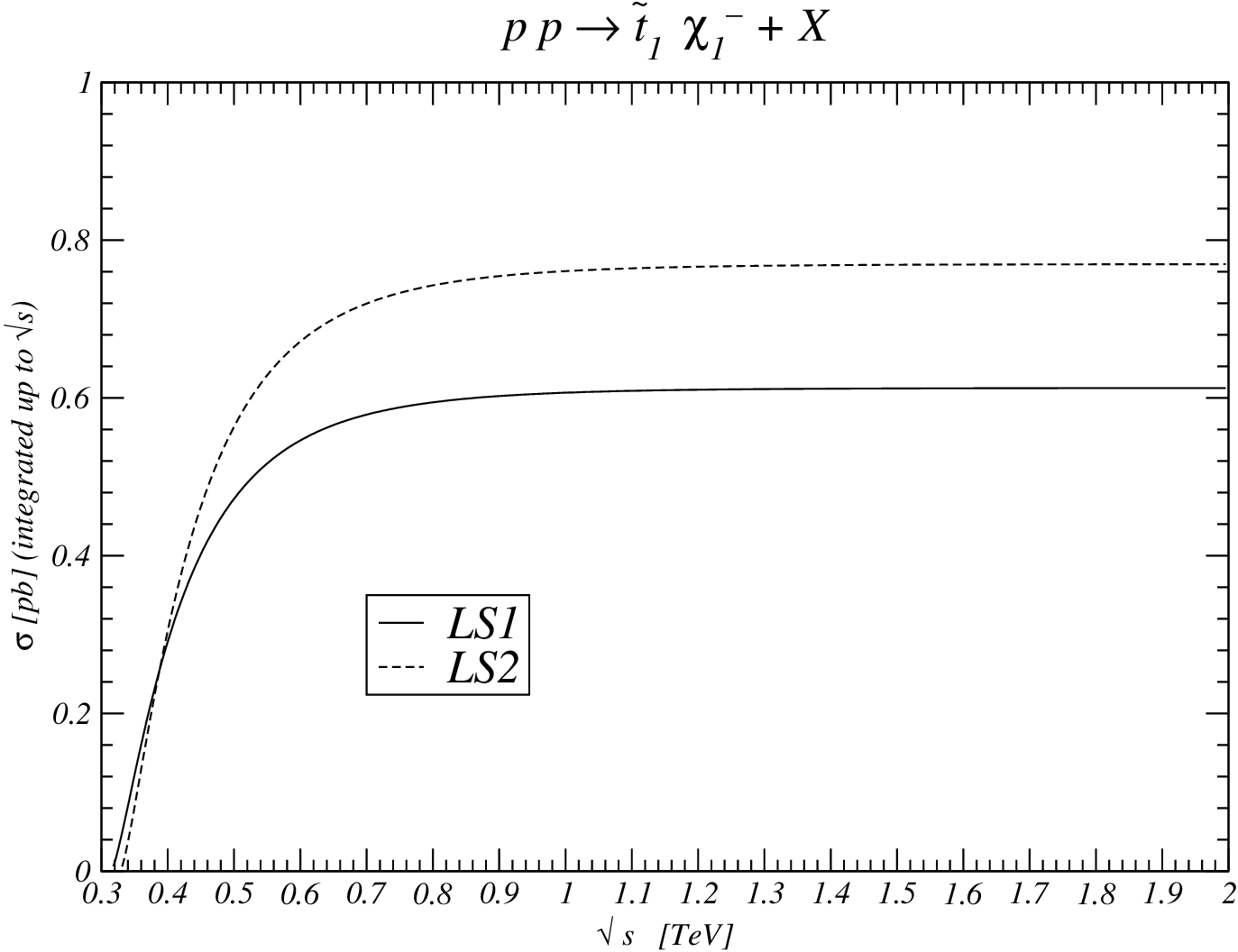, width=18cm, angle=90}
\vspace{1.5cm}
\caption{Integrated cross section from threshold up to $\sqrt{s}$ for the production of the  lightest  $\widetilde{t}\,\chi$ final state. Benchmark points: LS1, LS2.}
\label{fig:LS-sigma}
\end{figure}

\newpage
\begin{figure}[tb]
\centering
\epsfig{file=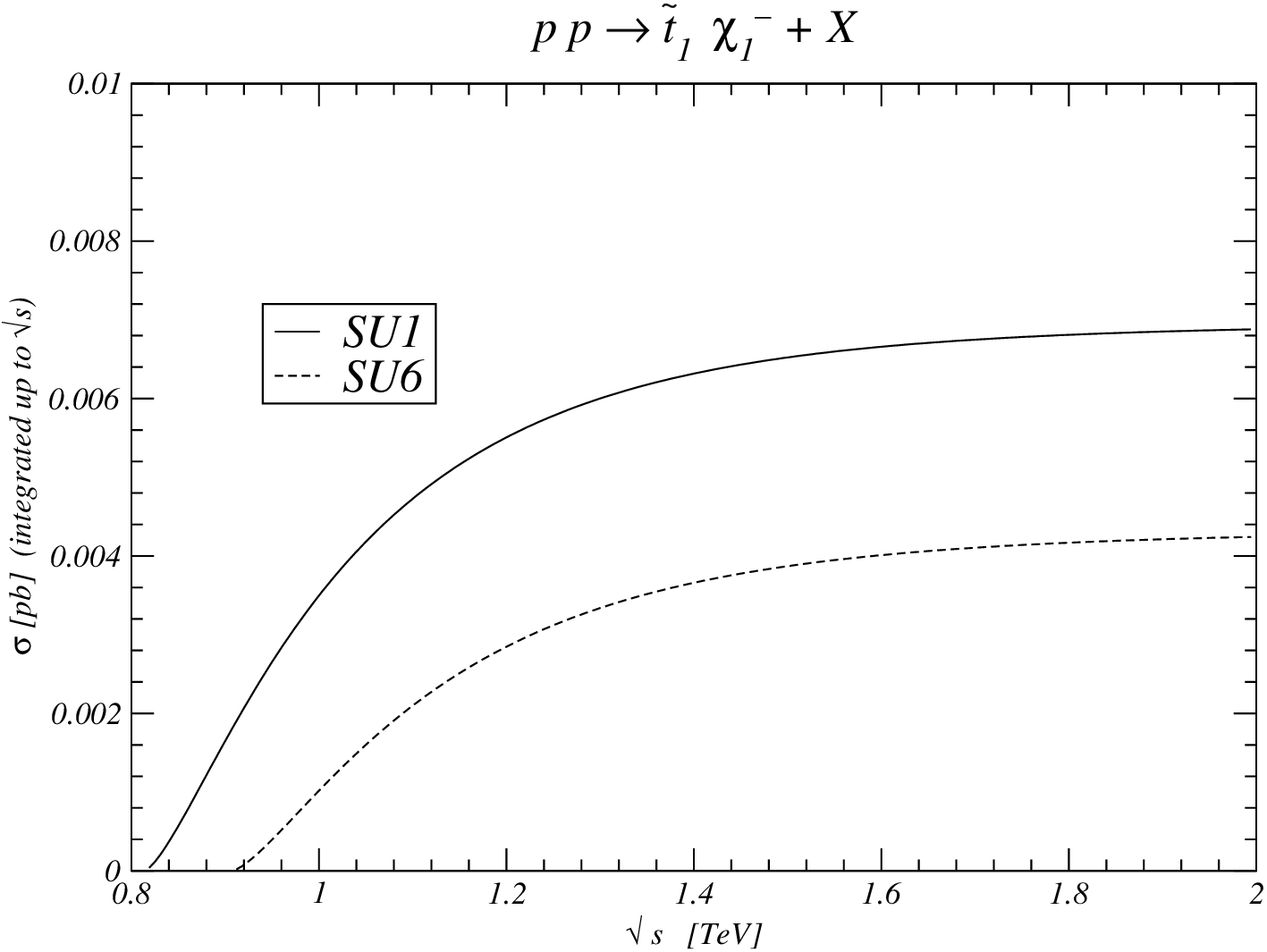, width=18cm, angle=90}
\vspace{1.5cm}
\caption{Integrated cross section from threshold up to $\sqrt{s}$ for the production of the  lightest  $\widetilde{t}\,\chi$ final state. Benchmark points: SU1, SU6.}
\label{fig:SU-sigma}
\end{figure}

\newpage
\begin{figure}[tb]
\centering
\epsfig{file=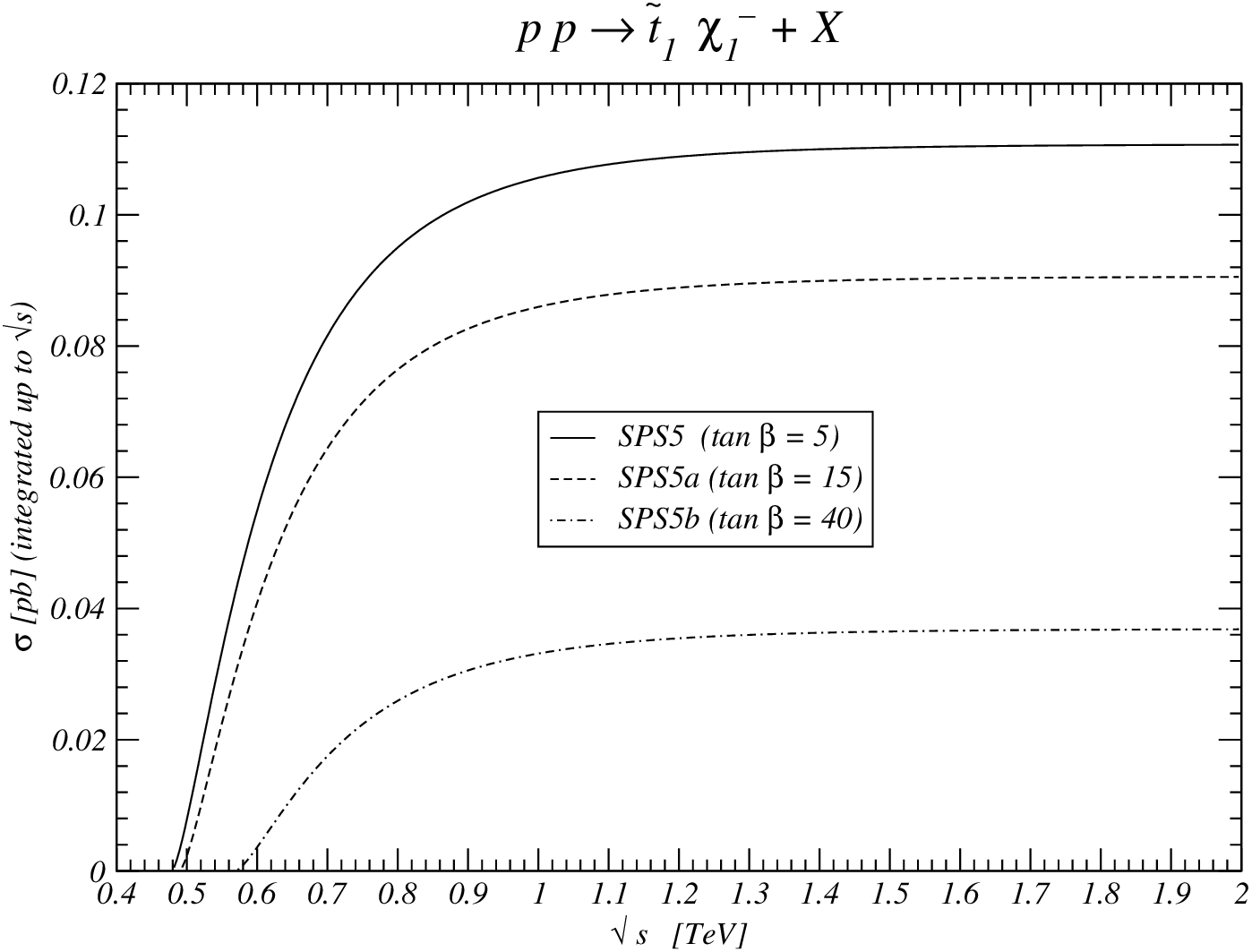, width=18cm, angle=90}
\vspace{1.5cm}
\caption{Integrated cross section from threshold up to $\sqrt{s}$ for the production of the  lightest  $\widetilde{t}\,\chi$ final state. Benchmark points: SPS5, SPS5a, SPS5b.}
\label{fig:SPS-sigma}
\end{figure}

\newpage

\newpage

\end{document}